\documentclass[runningheads]{llncs}
\usepackage{graphicx}
\usepackage{amsmath}
\usepackage{amssymb}
\usepackage{booktabs}
\usepackage{microtype}
\usepackage{times}
\usepackage{epsfig}
\usepackage{graphicx}
\usepackage{mathtools}
\usepackage{multirow}
\usepackage{tabularx}
\usepackage{array}
\usepackage{colortbl}
\usepackage{booktabs}
\usepackage{bbm}
\usepackage{multicol}
\usepackage{makecell}
\usepackage{xcolor}
\usepackage{xspace}
\usepackage{float}
\usepackage{color}
\usepackage{pifont}
\usepackage{hyperref}
\hypersetup{
    colorlinks=true,
    linkcolor=red,
    filecolor=red,      
    urlcolor=magenta,
}
\makeatletter
\newcommand{\printfnsymbol}[1]{%
	\textsuperscript{\@fnsymbol{#1}}%
}
\begin{document}
\title{ReconFormer: Accelerated MRI Reconstruction Using Recurrent Transformer}
%
\author{Pengfei Guo\thanks{Equal contribution.}\inst{1}, Yiqun Mei\printfnsymbol{1}\inst{1}, Jinyuan Zhou\inst{2}, Shanshan Jiang\inst{2}, Vishal M. Patel\inst{1,3}}
\authorrunning{Guo et al.}
%
\institute{Department of Computer Science, Johns Hopkins University, Baltimore, MD, USA\\
	\and
	Department of Radiology, Johns Hopkins University, Baltimore, MD, USA
	\and
	Department of Electrical and Computer Engineering, Johns Hopkins University, Baltimore, MD, USA\\
	}

\maketitle              
\begin{abstract}
Accelerating magnetic resonance image (MRI) reconstruction  process is a challenging ill-posed inverse problem due to the excessive under-sampling operation in $k$-space. 
In this paper, we propose a recurrent transformer model, namely \textbf{ReconFormer}, for MRI reconstruction which can iteratively reconstruct high fertility magnetic resonance images from highly under-sampled $k$-space data. In particular, the proposed architecture is built upon Recurrent Pyramid Transformer Layers (RPTL), which jointly exploits intrinsic multi-scale information at every architecture unit as well as the dependencies of the deep feature correlation through recurrent states. Moreover, the proposed ReconFormer is lightweight since it employs the recurrent structure for its parameter efficiency. We validate the effectiveness of ReconFormer on multiple datasets with different magnetic resonance sequences and show that it achieves significant improvements over the state-of-the-art methods with better parameter efficiency. Implementation code will be available in \href{https://github.com/guopengf/ReconFormer}{https://github.com/guopengf/ReconFormer}. 

\end{abstract}
\section{Introduction}
Magnetic Resonance Imaging (MRI) is one of the most prevalent diagnostic and research tools in clinical scenarios, which provides excellent resolution and abundant contrast mechanisms to visualize different structural and functional properties of the underlying anatomy. Due to physiological and hardware constraints~\cite{cs}, MRI acquisition process is inherently slow. Consequently, extending acquisition time to collect complete data in $k$-space (frequency domain) imposes significant burden on patients and makes MRI less accessible. One of the common approaches to accelerate MRI acquisition procedure is to collect partial data rather than traverse whole $k$-space, as known as under-sampling $k$-space. However, such operation violates the Nyquist-Shannon sampling theorem and introduces aliasing artifacts in the reconstructed image. Compressed sensing (CS) mitigates this issue by formulating the image reconstruction as solving an optimization problem with several assumptions including sparsity and incoherence~\cite{csmri}. Past literature in advanced CS-based image reconstruction has exploited low rank constraint terms~\cite{lowrank}, adaptive sparse modelling~\cite{dict,dict2}, and parallel imaging~\cite{parallel}. However, CS algorithms have several deficiencies impeding its practicability in real-world applications. First, CS recovery algorithms require careful tuning of regularization functions and hyper-parameters, which is also problem-specific and non-trivial. Second, due to the nature of the iterative optimization, CS methods often demand longer running time to achieve desirable reconstruction quality. Third, the acceleration factors are generally below 3 for typical MR images, when high-frequency oscillatory artifacts are not properly reduced during the optimization process~\cite{csdl}.

In contrast, recently advanced deep learning-based methods are gaining more attention for fast and accurate MRI reconstruction. Rather than explicitly defining prior information and regularization functions, networks learn them implicitly from data. CNN-based MRI reconstruction methods have been shown to provide much better MR image quality than conventional CS-based methods~\cite{crnn}. Such methods focus on elaborating architecture designs, including overcomplete networks~\cite{oucr}, neural ODEs~\cite{ode}, invertible networks~\cite{invert}, cascaded architectures~\cite{d5c5,kiki}, and recurrent structures~\cite{pcrnn,crnn}. Although
those methods achieve promising results, the basic convolution layer generally suffers from the following drawbacks: first, a content-independent convolution kernel is not optimal to restore different image regions~\cite{swinir}; second, the long-range dependency is not effectively modeled in convolutional networks~\cite{transformer}.
\begin{figure}[t]
\centering
\includegraphics[width=\textwidth]{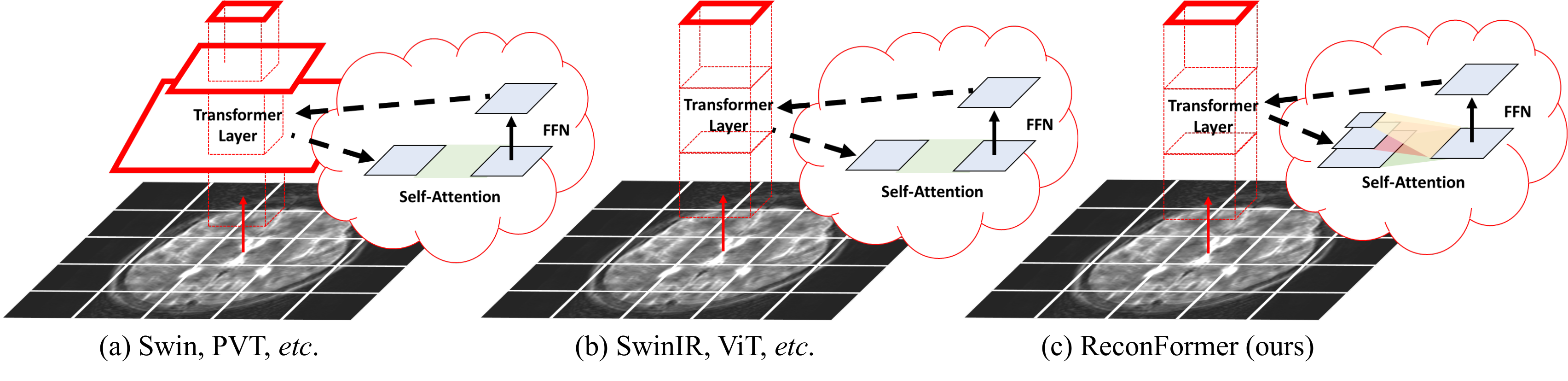}
\vspace{-9mm}
\caption{Comparisons of different transformer architectures.
(a) Recent high-level vision Transformer~\cite{swin,pvt} use a pyramid structure to model multi-scale contexts. (b) Previous low-level vision transformers~\cite{ipt,swinir} and ViT~\cite{vit} follow a simple columnar design, which preserves better information but fails to model at scales. (c) ReconFormer incorporates the pyramid structure inside a transformer layer. The local pyramid enables scale processing at each unit while the globally columnar structure maintains high-resolution information. FFN refers to feed-forward network in transformer layers.
}\label{fig:diff_arch}
\vspace{-7mm}
\end{figure}

Recent advances in transformer~\cite{transformer} models introduce the self-attention mechanism to capture global interactions between contexts and open a new possibility to solve the challenging problem of MRI reconstruction. While vision transformer has shown superior performance on general vision tasks~\cite{object,vit,swin,efficientvit}, a few attempts have been made for utilizing transformers in MRI reconstruction. Feng \emph{et al.}~\cite{TaskTransformer} proposed a task transformer network for joint MRI
reconstruction and super-resolution. Korkmaz \emph{et al.}~\cite{unsupervisedmrirecon} proposed SLATER, which can perform self-supervised MRI reconstruction facilitated by an adversarial transformer. A texture transformer module was introduced in~\cite{ttm} for reference-based MRI reconstruction. Although those works \cite{TaskTransformer,ttm,unsupervisedmrirecon} demonstrated a good performance, their architectures still heavily rely on the self-attention mechanism proposed in ViT~\cite{vit} and cannot efficiently model the multi-scale information. A fundamental vision principle that visual elements vary at scales  plays a vital role in computer vision~\cite{nonlocalrnn}. To leverage this intrinsic property, previous vision transformer models~\cite{object,vit,swin,efficientvit,TaskTransformer,unsupervisedmrirecon} rely on the network topology, as shown in Fig.~\ref{fig:diff_arch}(a). However, such design suffers from the following. 1) Scale modeling is inefficient. Scale information is gradually processed through network hierarchy, so the multi-scale representation is only obtained at the end of stages. 2) Scale modeling is inflexible. The network topology design is usually task-dependent and hard to generalize for other tasks.

In this paper, to overcome those issues, a novel Recurrent Transformer, termed \textbf{ReconFormer}, is proposed to recover the fully-sampled image
from the under-sampled $k$-space data in accelerated MRI reconstruction. In summary, the following are our key contributions: \textbf{1.} ReconFormer introduces a powerful locally pyramidal but globally columnar architecture, which can perceive multi-scale representation at any stage while well preserving image details, as shown in Fig.~\ref{fig:diff_arch}(c). \textbf{2.} Recurrent Pyramid Transformer Layers (RPTL) are proposed to allow scale modeling at every basic building units and exploit deep feature correlation through recurrent states. \textbf{3.} By fully employing the parameter efficiency of recurrent structure, ReconFormer allows the training on limited medical data from scratch. The effectiveness of our approach is validated on two datasets with various MR sequences and experiment results demonstrate that ReconFormer achieves significant improvements over the state-of-the-art CNN-based as well as transformer models with better parameter efficiency.

\section{Methodology}

\noindent {\bf{MRI Reconstruction. }} Let $x\in \mathbb{C}^N$ represent the observed complex-valued under-sampled $k$-space measurements and $y\in \mathbb{C}^M$ denote the fully-sampled data. Our aim is to reconstruct $y$ from $x$ ($N \ll M$) as follows:
\setlength{\belowdisplayskip}{0.1pt} \setlength{\belowdisplayshortskip}{0.1pt}
\setlength{\abovedisplayskip}{0.1pt} \setlength{\abovedisplayshortskip}{0.1pt}
\begin{equation} \label{eq1}
 x = F_{\text{u}}y+\epsilon,
\end{equation}
where $ F_{u}\in \mathbb{C}^{N\times M}$ denotes an under-sampled Fourier encoding matrix controlling the acceleration factor (AF). $\epsilon\in \mathbb{C}^N$ represents the additive acquisition noise. $F$ and $F^{-1}$ represent the Fourier transform and its inverse. The MRI reconstruction is commonly formulated as an unconstrained optimization problem as follows~\cite{crnn,d5c5,pcrnn}:
\begin{equation}\label{eq2}
	\min\limits_{\text{y}} R(y) + \lambda \|x - F_{\text{u}}y\|_2^2.
\end{equation}
Here, $R$ is the regularization term on $y$ and $\lambda$ adjusts the contribution of the data fidelity term  based on the noise level of the acquired
measurements $x$. For deep learning-based approaches~\cite{crnn,d5c5}, one can force $y$ to be approximated in the regularization term $R$ by reconstruction. The optimization problem in Eq.~\ref{eq2} can be rewritten as follows:
\begin{equation}\label{eq3}
	\min\limits_{\text{y}} \|y - F[f_{\text{nn}}(\bar{x}|\Theta)] \|_2^2 + \lambda \|x - F_{\text{u}}y\|_2^2,
\end{equation}
where $f_{\text{nn}}$ denotes the forward pass of a neural network parameterized by $\Theta$ and $\bar{x}$ is the zero-filled reconstruction. We denote $\bar{y}=f_{\text{nn}}(\bar{x}|\Theta)$ as the approximated fully-sampled reconstructed image from the observed under-sampled $k$-space data $x$. 

\begin{figure}[ht!]
\centering
\includegraphics[width=\textwidth]{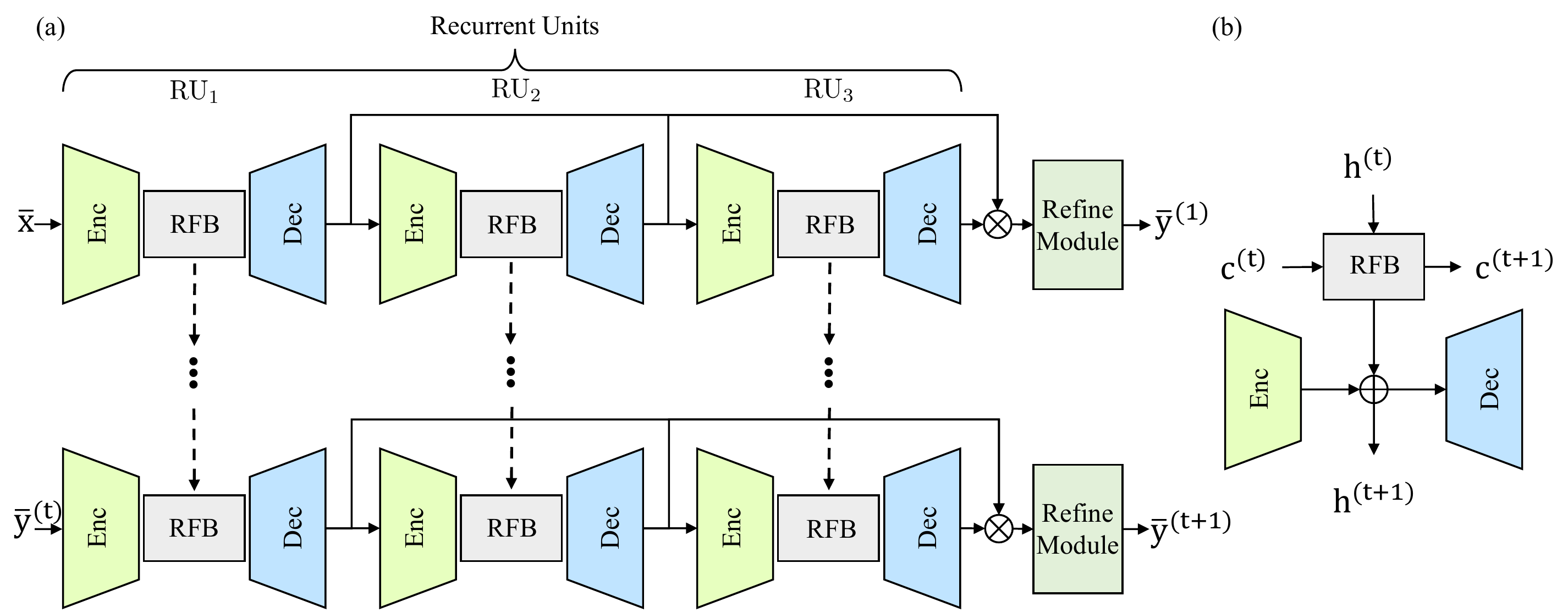}
\vspace{-7mm}
\caption{The architecture of the proposed ReconFormer.
(a) A schematic of unrolled ReconFormer iterations. Here, $\otimes$ denotes the channel-wise concatenation and RFB is the ReconFormer block.  (b) The illustration of a recurrent unit (RU). Here, $\oplus$ denotes the element-wise addition.
}\label{fig:pipline}
\vspace{-8mm}
\end{figure}
\noindent {\bf{ReconFormer. }} As shown in Fig.~\ref{fig:pipline}(a), ReconFormer consists of three recurrent units and a refine module (RM). To maintain high-resolution information, ReconFormer employs globally columnar structure. In particular, recurrent units map the input degraded images to the original dimension. Meanwhile, across each recurrent unit, the receptive fields of recurrent units is gradually reduced to reconstruct high-quality MR images in a coarse-to-fine way. It is worth noting that the last recurrent unit $\text{RU}_3$ employs overcomplete architecture~\cite{oucr}, which has been demonstrated to efficiently constraint the receptive field. As shown in Fig.~\ref{fig:pipline}(b), a recurrent unit contains an encoder $f^{\text{Enc}}$, a ReconFormer block $f^{\text{RFB}}$, and a decoder $f^{\text{Dec}}$. To have a more stable optimization~\cite{conv_trans}, encoder and decoder are built up on convolution layers. A data consistency (DC) layer is added at the end of each decoder network to reinforce the data consistency in the $k$-space. We denote a single recurrent unit $\text{RU}_i$ as $f^{\text{RU}}_i$ where $i\in\{1,2,3\}$ indicates the $i^{\text{th}}$ recurrent unit and its operation in the $t^{\text{th}}$ iteration can be formulated as follows:
\begin{equation} \label{eq4}
\begin{aligned} 
\bar{y}_{i}^{(t)} &= \text{DC}(f^{\text{RU}}_{i}(\bar{y}_{i-1}^{(t)},h_i^{(t)}, c_i^{(t)}),x,U), \\
 &= F^{-1}[Ux+(1-U)F[(f^{\text{RU}}_{i}(\bar{y}_{i-1}^{(t)},h_i^{(t)},c_i^{(t)})]],\\
 &= F^{-1}[Ux+(1-U)F[f_{i}^{\text{Dec}}(f_i^{\text{RFB}}(h_i^{(t)},c_i^{(t)}) + f_{i}^{\text{Enc}}(\bar{y}_{i-1}^{(t)}) ) ]],\\
\end{aligned}
\end{equation}
where $U$ is a binary under-sampling mask in $F_{\text{u}}$ which is used for data consistency in $k$-space. $\{\bar{y}_{i-1}^{(t)},\bar{y}_{i}^{(t)}\}\in \mathbb{R}^{H\times W \times 2}$ are input and output of $\text{RU}_i$, respectively, where $H$ and $W$ are the zero-filled reconstruction height and width, and 2 represents the real and imaginary channels. $h_i^{(t)} \in \mathbb{R}^{\frac{H}{S} \times \frac{W}{S} \times C}$ is the hidden state from the previous iteration, where $S$ is the scaling factor determined by the structure of each RU and consequently controls the size of receptive fields. $c_i^{(t)}$ is deep feature correlation from the previous iteration. At the $t^{\text{th}}$ iteration, ReconFormer can be unrolled as follows:
\begin{equation} \label{eq5}
\begin{aligned} 
\{\bar{y}_{1}^{(t)},\bar{y}_{2}^{(t)},\bar{y}_{3}^{(t)}\} &=\{ \text{RU}_i(\bar{y}_{i-1}^{(t)},h_i^{(t)}, c_i^{(t)},x,U) | i =1,2,3\} ,\\
\bar{y}^{(t+1)} &= \text{DC}(\text{RM}(\bar{y}_{1}^{(t)} \otimes \bar{y}_{2}^{(t)} \otimes \bar{y}_{3}^{(t)}), x, U),\\
\end{aligned}
\end{equation}
where $\otimes$ denotes the channel-wise concatenation. 
$\bar{y}^{(t+1)}$ is the output of the current iteration and also servers as the input for the next iteration. 

\begin{figure}[t]
\centering
\includegraphics[width=0.75\textwidth]{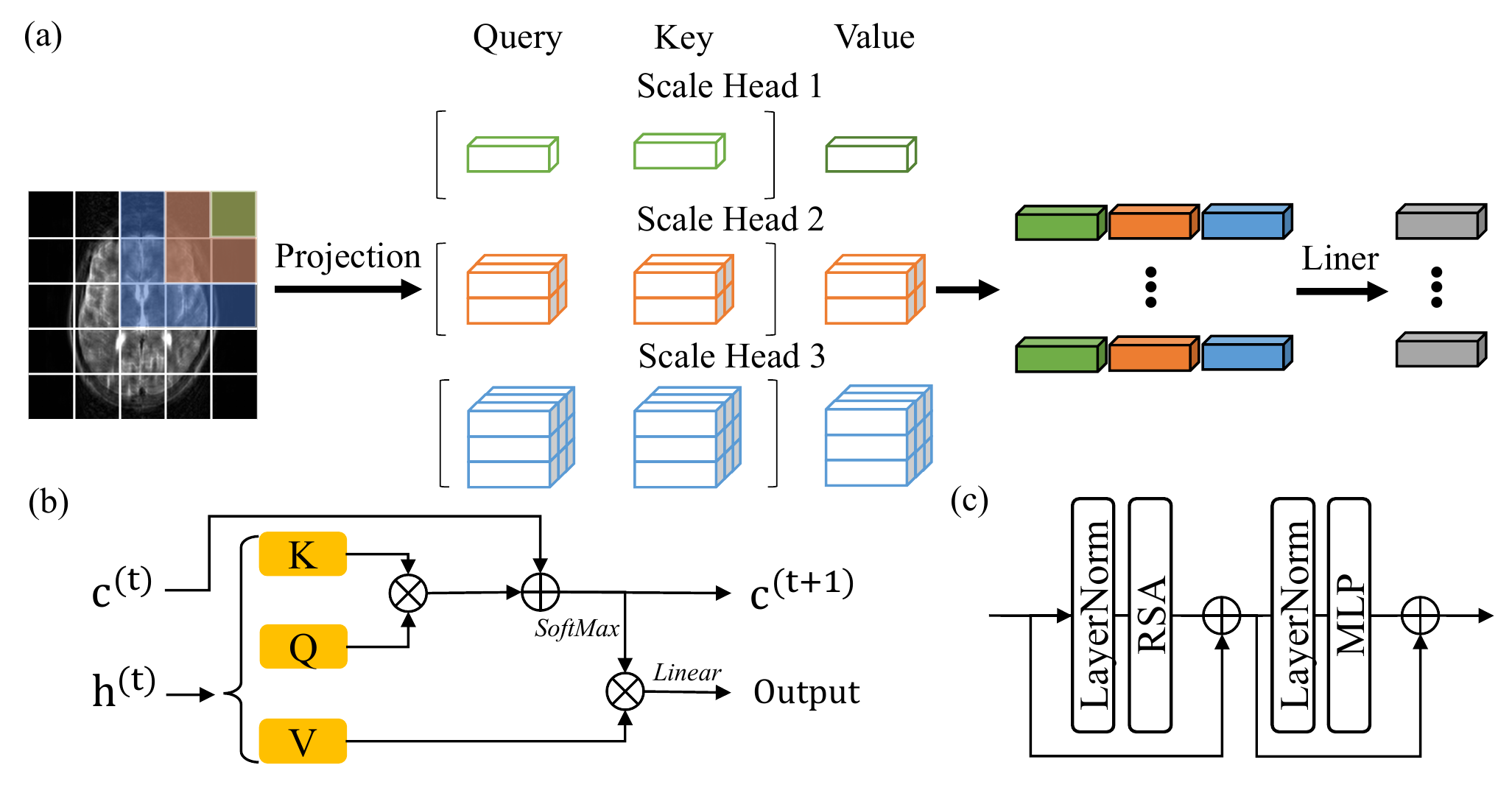}
\vspace{-5mm}
\caption{ (a) The schematic of recurrent scale-wise attention (RSA). (b) An illustration of the transition function in a RPTL. (c) The architecture of the proposed RPTL.
}\label{fig:rptl}
\vspace{-6.0mm}
\end{figure}
\noindent {\bf{RPTL. }} A ReconFormer block (RFB) is formed by stacked recurrent pyramid transformer layers (RPTL). The core design of RPTL is the Recurrent Scale-wise Attention (RSA), as shown in Fig.~\ref{fig:rptl}(a). Compared to the standard multi-head self-attention of the original transformer layer~\cite{vit}, the main
difference lies in two aspects: (i) instead of performing a single attention function, the proposed RSA consists of several attention scale heads (\emph{i.e.}, $\times1, \times3, \times5 $), which operate on multi-scale patches in parallel. Such design enables efficient in-place scale modeling and forms a feature pyramid by projecting features at various scales directly into multiple attention heads. Consequently, the proposed RPTL allows scale processing at basic architecture unit. (ii) The correlation estimation in the proposed RSA relies on both the hidden state $h^{(t)}$ and the deep feature correlation $c^{(t)}$ from the previous iteration, which enables more robust correlation estimation by propagating  correlation information between adjacent states. The proposed RPTL utilizes local attention mechanism with shifted window scheme~\cite{swin}. Given the input hidden state $h^{(t)} \in \mathbb{R}^{H^\prime \times W^\prime \times C^\prime }$, RPTL first uses non-overlapping $K \times K$ local windows to partition $h^{(t)}$ into a feature of size $\frac{H^\prime \times W^\prime }{K^2} \times K^2 \times C^\prime $, where $K$ is the size of local windows and $\frac{H^\prime \times W^\prime }{K^2}$ is the number of windows. Then, the proposed RSA operates on each window in parallel to compute the self-attention. Let $F \in \mathbb{R}^{ K^2 \times C^\prime}$ denote a local window feature. the query $Q$,
key $K$ and value $V$ matrices in RPTL are formulated as follows:
\begin{equation} \label{eq6}
\begin{aligned} 
Q = F P_Q, K = F P_K, V = F P_V,
\end{aligned}
\end{equation}
where $P_Q, P_K,\text{and } P_V$ are the  corresponding projection matrices in attention scale heads. As shown in Fig.~\ref{fig:rptl}(b), we compute the outputs of RSA as follows:
\begin{equation} \label{eq7}
\begin{aligned} 
\text{Attention}(Q,K,V,c^{(t)}) &= \text{SoftMax}(c^{(t+1)})V ,\\
c^{(t+1)} &= \lambda\frac{QK^T}{\sqrt{d}} + (1-\lambda)c^{(t)},
\end{aligned}
\end{equation}
where $d$ is the embedding dimension and $\lambda$ is a learnable parameter controlling the contribution of the deep feature correlation from the adjacent state. As shown in Fig.~\ref{fig:rptl}(c), we can formulate the whole process of RPTL as follows:
\begin{equation} \label{eq8}
\begin{aligned} 
F &= \text{RSA}(\text{LN}(F))+F,\\
F &= \text{MLP}(\text{LN}(F))+F,
\end{aligned}
\end{equation}
where LN and MLP denote the Layer Normalization and the multi-layer perceptron, respectively. The detailed network configuration can be found in the supplementary material.
\begin{figure}[ht!]
\centering
\includegraphics[width=\textwidth]{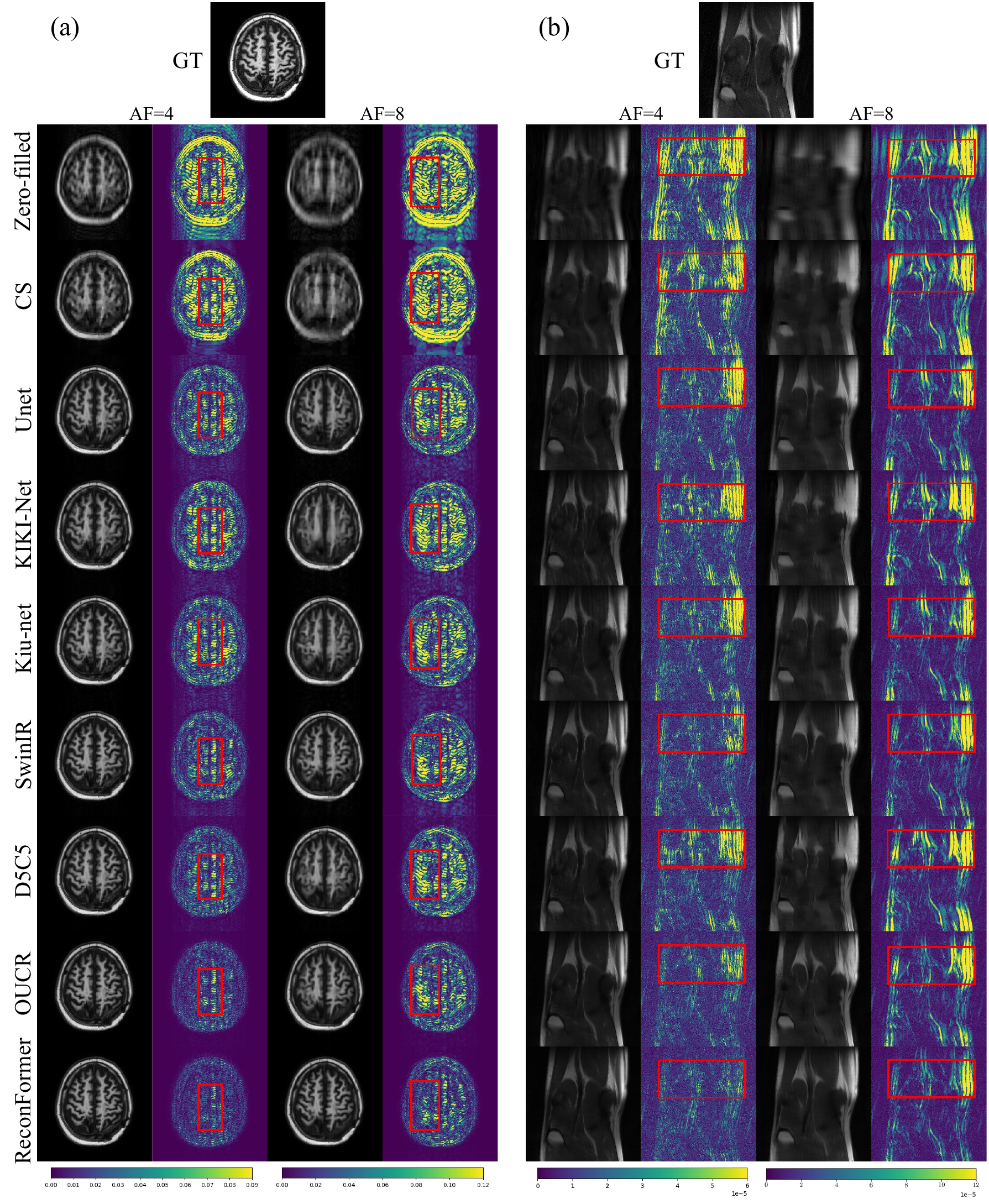}
\vspace{-9mm}
\caption{Qualitative comparison of different methods on (a) the HPKS dataset and (b) the fastMRI dataset. The second column of each subplot shows the corresponding error maps. The red boxes highlight regions where RenconFormer significantly outperforms the competitive methods.
}\label{fig:vis}
\vspace{-7mm}
\end{figure}

\section{Experiments and Results}
\noindent {\bf{Evaluation and Implementation Details. }} The \textbf{fastMRI}~\cite{fastmri} and \textbf{HPKS}~\cite{hpks} datasets are used for conducting experiments. The fastMRI dataset contains 1,172 complex-valued single-coil coronal proton density (PD)-weighted knee MRI scans. Each scan approximately provides 35 coronal cross-sectional knee images with the matrix of size $320 \times 320$. We partition this dataset into 973 scans for training, and 199 scans (fastMRI validation dataset) for testing. 
The HPKS dataset provides complex-valued single-coil axial $T_1$-weighted brain MRI scans from 144 post-treatment patients with malignant glioma. Each scan contains 15 axial cross-sectional images with the matrix of size $256 \times 256$. The data splits are as follows: 102 scans are used for training, 14 scans are used for validation, and 28 scans are used for testing. In experiments, the input under-sampled image sequences are generated by randomly under-sampling the $k$-space data using the Cartesian under-sampling function that is the same as the fastMRI challenge~\cite{fastmri}. The proposed ReconFormer is trained using the $\ell_{1}$ loss with Adam optimizer with the following hyperparameters: learning rate of $2.0 \times 10^{-4}$; 50 maximum epochs; batch size of 4; the number of recurrent iterations $T$ of 5.  Peak signal to noise ratio (PSNR) and structural index similarity (SSIM) are used as the evaluation metrics for comparison. We implement the proposed model using PyTorch on Nvidia RTX8000 GPUs.
\begin{figure}[t!]
\centering
\includegraphics[width=\textwidth]{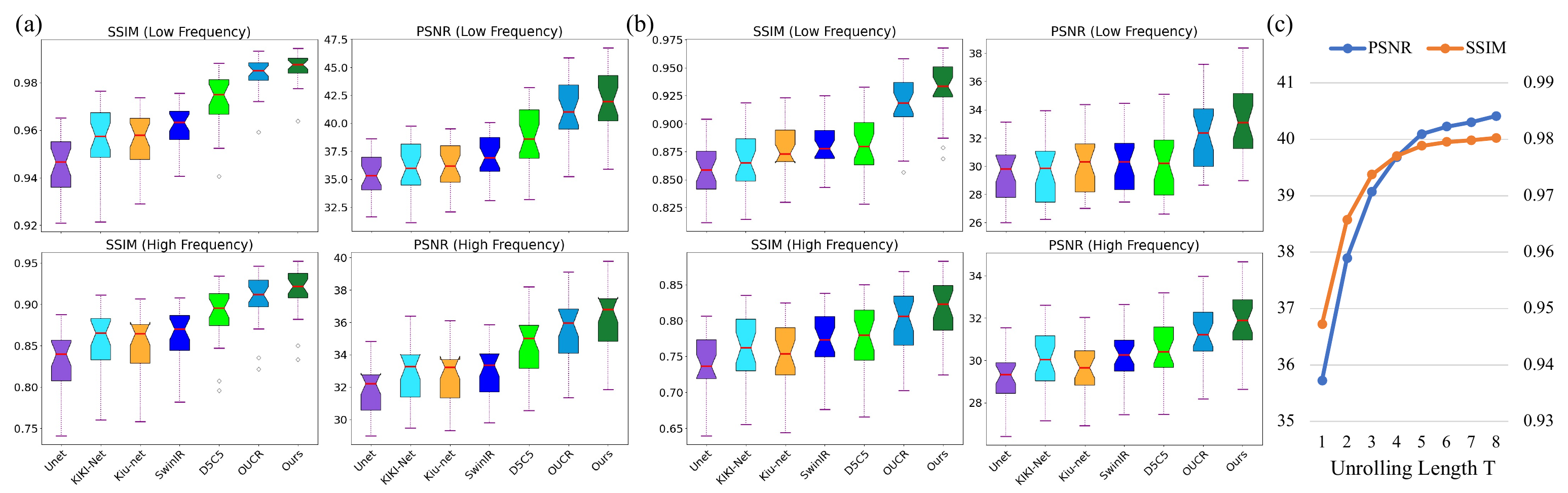}
\vspace{-9mm}
\caption{$k$-space analysis of (a) low frequency and (b) high frequency on the HPKS dataset. (c) Unrolling length v.s. reconstruction performance of ReconFormer on HPKS dataset with $4\times$ acceleration.
}\label{fig:kspace}
\vspace{-2mm}
\end{figure}

\begin{table}[t!]
	\centering
	\setlength{\tabcolsep}{5pt}
	\scriptsize
	\caption{Quantitative results on the HPKS and fastMRI dataset. Param denotes the number of parameters.}\label{tab1}
	\vskip-3mm
\begin{tabular}{l|c|cccc|cccc}
\hline
\multirow{3}{*}{Method} & \multirow{3}{*}{Param} & \multicolumn{4}{c|}{HPKS}                                                                                                          & \multicolumn{4}{c}{fastMRI}                                                                                                        \\ \cline{3-10} 
                        &                        & \multicolumn{2}{c|}{SSIM}                                                   & \multicolumn{2}{c|}{PSNR}                            & \multicolumn{2}{c|}{SSIM}                                                   & \multicolumn{2}{c}{PSNR}                             \\ \cline{3-10} 
                        &                        & \multicolumn{1}{c|}{AF = 4}              & \multicolumn{1}{c|}{AF = 8}              & \multicolumn{1}{c|}{AF = 4}             & AF = 8            & \multicolumn{1}{c|}{AF = 4}              & \multicolumn{1}{c|}{AF = 8}              & \multicolumn{1}{c|}{AF = 4}             & AF = 8            \\ \hline
CS~\cite{cs_recon}      & -                      & \multicolumn{1}{c|}{0.8705}          & \multicolumn{1}{c|}{0.7125}          & \multicolumn{1}{c|}{29.94}          & 24.96          & \multicolumn{1}{c|}{0.5736}          & \multicolumn{1}{c|}{0.4870}          & \multicolumn{1}{c|}{29.54}          & 26.99          \\ \hline
UNet~\cite{unet}        & 8.634 M                & \multicolumn{1}{c|}{0.9155}          & \multicolumn{1}{c|}{0.8249}          & \multicolumn{1}{c|}{34.47}          & 29.47          & \multicolumn{1}{c|}{0.7142}          & \multicolumn{1}{c|}{0.6424}          & \multicolumn{1}{c|}{31.88}          & 29.78          \\ \hline
KIKI-Net~\cite{kiki}    & 1.790 M                & \multicolumn{1}{c|}{0.9363}          & \multicolumn{1}{c|}{0.8436}          & \multicolumn{1}{c|}{35.35}          & 29.86          & \multicolumn{1}{c|}{0.7172}          & \multicolumn{1}{c|}{0.6355}          & \multicolumn{1}{c|}{31.87}          & 29.27          \\ \hline
Kiu-net~\cite{kiunet}   & 7.951 M                & \multicolumn{1}{c|}{0.9335}          & \multicolumn{1}{c|}{0.8467}          & \multicolumn{1}{c|}{35.35}          & 30.18          & \multicolumn{1}{c|}{0.7228}          & \multicolumn{1}{c|}{0.6456}          & \multicolumn{1}{c|}{32.06}          & 29.86          \\ \hline
SwinIR~\cite{swinir}    & 5.156 M                & \multicolumn{1}{c|}{0.9364}          & \multicolumn{1}{c|}{0.8499}          & \multicolumn{1}{c|}{36.00}          & 30.42          & \multicolumn{1}{c|}{0.7213}          & \multicolumn{1}{c|}{0.6537}          & \multicolumn{1}{c|}{32.14}          & 30.21          \\ \hline
D5C5~\cite{d5c5}        & 2.237 M                & \multicolumn{1}{c|}{0.9595}          & \multicolumn{1}{c|}{0.8623}          & \multicolumn{1}{c|}{37.51}          & 30.40          & \multicolumn{1}{c|}{0.7256}          & \multicolumn{1}{c|}{0.6457}          & \multicolumn{1}{c|}{32.25}          & 29.65          \\ \hline
OUCR~\cite{oucr}        & 1.192 M                & \multicolumn{1}{c|}{0.9747}          & \multicolumn{1}{c|}{0.9044}          & \multicolumn{1}{c|}{39.33}          & 32.14          & \multicolumn{1}{c|}{0.7354}          & \multicolumn{1}{c|}{0.6634}          & \multicolumn{1}{c|}{32.61}          & 30.59          \\ \hline
ReconFormer (Ours)            & \textbf{1.141 M}       & \multicolumn{1}{c|}{\textbf{0.9788}} & \multicolumn{1}{c|}{\textbf{0.9210}} & \multicolumn{1}{c|}{\textbf{40.09}} & \textbf{33.04} & \multicolumn{1}{c|}{\textbf{0.7383}} & \multicolumn{1}{c|}{\textbf{0.6697}} & \multicolumn{1}{c|}{\textbf{32.73}} & \textbf{30.89} \\ \hline
\end{tabular}
\vskip-6mm
\end{table}

\noindent {\bf{Comparisons with the state-of-the-art. }}
To verify the effectiveness of the proposed ReconFormer,
we compare the proposed method with 7 representative methods, including conventional compressed sensing (CS) based method~\cite{cs_recon}, popular CNN-based methods -- UNet~\cite{unet},  KIKI-Net~\cite{kiki}, Kiu-net~\cite{kiunet}, and D5C5~\cite{d5c5}, state-of-the-art iterative reconstruction approaches -- OUCR~\cite{oucr}, and vision transformer model -- SwinIR~\cite{swinir}. For a fair comparison, methods (UNet~\cite{unet}, Kiu-net~\cite{kiunet}, and SwinIR~\cite{swinir}) that are not originally proposed for MRI reconstruction are modified for data with real
and imaginary channels and a DC layer is added at the end of the networks. The unrolling length of all iterative approaches is set to 5.

Table~\ref{tab1} shows the quantitative results evaluated on the two datasets for $\text{AF=4}$ and $\text{AF=8}$. Compared with the other methods, the proposed ReconFormer achieves the best performance on multiple datasets for all acceleration factors while containing the least number of parameters. It is worth noting that our method exhibits a larger performance improvement, when the acceleration factor increases (\emph{i.e.},  more challenging scenarios). In particular, for the HPKS and fastMRI datasets, our model outperforms the most competitive approach OUCR~\cite{oucr}
by 0.9 dB and 0.3 dB in 8$\times$ acceleration, respectively. While the fastMRI dataset is more challenging due to the acquisition quality, all reported improvements achieved by ReconFormer are statistically significant. We provide the detailed statistical significance test in the supplementary material. 
Figure~\ref{fig:vis} shows qualitative evaluations on two datasets. It can be seen that ReconFormer yields remarkable reconstruction quality and perceptually outperforms the other methods by a large margin, as shown in the red boxes of each sub-figure in Fig.~\ref{fig:vis}. Moreover, we conduct the $k$-space analysis to investigate the performance of different methods in low (1/3 center frequencies) and high (2/3 peripheral frequencies) spatial frequencies. As shown in Fig.~\ref{fig:kspace}, the proposed method is shown to be more effective for both low frequency information (\emph{e.g.}, contrast, brightness, and general shape) and high frequency features (\emph{e.g.}, edges, details, sharp transitions).
The superior performance on different MRI sequences demonstrates the merit of the proposed transformer architecture and our RPTL in jointly utilizing intrinsic multi-scale information and the deep feature correlation. 

\noindent {\bf{Ablation Study. }} We first separately evaluate the effectiveness of network components, including three recurrent units, RM, and PRTL in the proposed ReconFormer. As shown in Table~\ref{tab2}, when we add each recurrent unit, the reconstruction quality can be gradually boosted, which demonstrates the effectiveness of the architecture design. Then, we show that adding RM to fuse the coarse-to-fine reconstructions further improves the performance. Moreover, RPTL consistently provides performance improvements in different acceleration factors, which implies the importance of fully exploiting multi-scale information in basic building blocks. For experiments without RPTL, we replace RSA by a standard multi-head self attention. The influence of the unrolling length T on reconstruction results is shown in Fig.~\ref{fig:kspace}(c). We can observe that  reconstruction quality increases as the unrolling length expands, but gets saturated when T = 5. There is a trade-off between the reconstruction accuracy and the computational efficiency. Due to the space limitation, computation details of different methods, more visualizations, and analysis on the fastMRI dataset are provided in the supplementary material.

\begin{table}[t!]
\vskip-8pt
	\setlength{\tabcolsep}{8.0pt}
	\scriptsize
	\centering
\caption{The Ablation study of proposed modules on HPKS.}\label{tab2}
\vskip-4mm
\begin{tabular}{ccccc|cc|cc}
\hline
\multicolumn{5}{c|}{Modules}                                   & \multicolumn{2}{c|}{AF=4} & \multicolumn{2}{c}{AF=8} \\ \hline
RU$_1$     & RU$_2$     & RU$_3$     & RM         & RPTL       & SSIM         & PSNR       & SSIM        & PSNR       \\ \hline
\checkmark &            &            &            &            & 0.9671       & 38.19      & 0.8900      & 31.39      \\
\checkmark & \checkmark &            &            &            & 0.9762       & 39.63      & 0.9105      & 32.47      \\
\checkmark & \checkmark & \checkmark &            &            & 0.9773       & 39.79      & 0.9151      & 32.75      \\
\checkmark & \checkmark & \checkmark & \checkmark &            & 0.9783       & 39.94      & 0.9186      & 32.90      \\
\checkmark & \checkmark & \checkmark & \checkmark & \checkmark & 0.9788       & 40.09      & 0.9210      & 33.04      \\ \hline
\end{tabular}
\vskip-8mm
\end{table}
\section{Conclusion}
In this paper, we propose a recurrent transformer-based MRI reconstruction model ReconFormer. By leveraging the novel RPTL, we are able to explore the multi-scale representation at every basic building units and discover the dependencies of the deep feature correlation between adjacent recurrent states. Incorporating three recurrent units and the refine module, ReconFormer reconstructs high-quality MR images through a locally pyramidal but globally columnar structure and achieves the state-of-the-art performance on multiple datasets.  ReconFormer is lightweight and does not require pre-training on large-scale datasets. Our experiments suggest the promising potential of using transformer-based models in the MRI reconstruction task. In the future, we will extend the proposed model to multi-coil reconstruction and more MRI sequences.

\bibliographystyle{splncs04}
\bibliography{references}

\end{document}